\documentclass[aps,pra,showpacs,groupedaddress]{revtex4}
\usepackage{amsmath}
\usepackage{mathrsfs}
\usepackage{amssymb}
\usepackage[dvips]{graphics,color}
\usepackage{graphicx}
\usepackage{subfigure}
\usepackage{dcolumn}
\usepackage{bm}

\begin{document}

\title{The improved Gaussian approximation Calculation of Bogoliubov Mode in
One Dimensional Bosonic Gas}
\author{Qiong Li}
\affiliation{Department of Physics, Peking University, Beijing, 100871, China}
\author{Daoguang Tu}
\affiliation{Department of Physics, Peking University, Beijing, 100871, China}
\author{Dingping Li}
\affiliation{Department of Physics, Peking University, Beijing, 100871, China}

\begin{abstract}
In this paper, we study the homogeneous one-dimensional bosonic gas
interacting via a repulsive contact potential by using the improved Gaussian
approximation. We obtain the gapless excitation spectrum of Bogoliubov mode.
Our result is in good agreement with the exact numerical calculation based
on the Bethe ansatz. We speculate that the improved Gaussian approximation
could be a quantitatively good approximation for higher dimensional systems.
\end{abstract}

\maketitle

\section{Introduction}

Since the concept of Bose-Einstein condensation (BEC) was originally put
forward by Bose and Einstein, the dilute Bose gas, as a many-body system
which displays macroscopic quantum phenomena such as superfluidity, has been
extensively studied theoretically. The microscopic description of BEC
started with Bogoliubov theory\cite{key-1,key-2,key-3,key-4}, in which the
destruction and creation operators for the macroscopically-occupied
lowest-energy mode is specially treated as c numbers, known as Bogoliubov
replacement. Based on Bogoliubov replacement, the Green's function methods
were applied to a dilute Bose gas at zero temperature \cite%
{key-5,key-6,key-7}. Hugenholtz and Pines \cite{key-7} showed that for a
repulsive interaction, the pole of the one-particle Green's function,
approaches zero for zero momentum, which means a gapless excitation spectrum
(usually we call it as Goldstone theorem \cite{key-41}). P.C. Hohenberg and
P.C. Martin described BEC as spontaneous global $U(1)$ symmetry breaking by
introducing external sources, which are set negligibly small in the end \cite%
{key-8}. The interpretation of BEC as symmetry breaking makes the quantum
field-theoretic treatment very convenient, in which the expectation value of
the field operator describes the density as well as the wavefunction of the
condensed bosons and hence is also called "macroscopic wavefunction". The
effective action approach \cite{key-33,Baym,key-34,key-35,key-36} is usually
employed and kinds of approximations can be easily formulated in this
framework, such as Bogoliubov approximation, Popov approximation and
Hartree-Fock Bogoliubov (HFB) approximation as discussed in detail in the
references \cite{key-9,key-10,key-11,key-12,key-13,key-14}.

However for the bosonic model, if we use the simplest non-perturbative
calculation, Hartree-Fock Bogoliubov (HFB) approximation, the spectrum
obtained is gapped even in the broken phase. Goldstone theorem is violated
in such approximation \cite{key-8,key-9}. Though in the Popov approximation,
the spectrum remains gapless, the method is not self-consistent and we will
also show that we can not apply this method to one dimensional bosonic
model. $\ \Phi $ derivable theory, self-consistent approximation method
beyond HFB, including some higher two particle irreducible (2PI) diagrams to
the effective action, is often used in studying BEC systems. The spectrum
obtained in the $\Phi $ derivable theory is also gapped \cite{Hendrik}.

In the self-consistent theories such as Hartree-Fock Bogoliubov (HFB)
approximation, the Ward identity from $U(1)$ symmetry is not preserved due
to partial resummations of some Feymann diagrams. Therefore, the Goldstone
theorem is violated and the resulting excitation spectrum is gapped even in
the symmetry breaking phase. \ In order to preserve the Ward identity, we
should incorporate the contributions of some other Feymann diagrams and
thereby remove the gap \cite{Rosenstein,Okopinska,Hendrik}. It is called
"covariant Gaussian approximation" in \cite{Rosenstein} and we will call it
"improved Gaussian approximation" (IGA). In principle we can apply similar
method to the $\Phi $ derivable theory beyond HFB (we will call it the
improved $\Phi $ derivable theory, or IDT in short), but the theory becomes
too complex (involving integral equations which can not be solved
analytically) \cite{Hendrik}.

In recent years, interest in 1D Bose gas has been revived due to its
experimental realization with ultracold bosonic atoms \cite%
{key-22,key-23,key-24,key-25}. In one dimension (1D), at finite temperature,
the excitation spectra are gapped. However, the 1D Bose gas at zero
temperature contains gapless spectra and the system is algebraic long range
order. In a trapped 1D gas, the Bose-Einstein condensation (BEC) regimes of
a true condensate, quasicondensate regime and the regime of a trapped Tonks
gas (gas of impenetrable bosons) at finite temperature have been identified
in \cite{ShlyapnikovPRL2000}. The stability and phase coherence of trapped
1D Bose gases was studied in \cite{ShlyapnikovPRL2003}. Most of the other
relevant works are summarized in the review article \cite{key-26}. In highly
anisotropic traps, where the axial motion of the atoms is weakly confined
while the radial motion is frozen by the tight transverse confinement, the
shape of the Bose-condensed systems reduces to one dimension. If the
characteristic range of the interatomic potential is much smaller than the
typical length of the radial extension, the system can be described by the
Lieb-Liniger model \cite{key-19,key-20}, in which the contact potential
strength $g_{1D}$ is given by $g_{1D}=-\frac{2\hbar ^{2}}{ma_{1D}}$ with $%
a_{1D}$ being the 1D scattering length\cite{key-27,key-28}. The Lieb-Liniger
model can be exactly solved by the Bethe ansatz and two types of excitations
(named Type I and Type II) have been found. Type I excitations are gapless
with a linear dispersion in the long wavelength limit and reduce to the
Bogoliubov excitations in the weak coupling limit. Type II excitations, the
Fermionic excitations which are prominent in the strong coupling regime,
have no equivalent in the Bogoliubov theory. J. S. Caux \textit{et al.}\cite%
{key-29} studied the one-particle dynamical correlation function of the
Lieb--{}Liniger model by using the ABACUS method \cite{key-30}, for a wide
range of values of the interaction parameter.

In this paper we will apply IGA to the 1D Bose gas at zero temperature. This
system can be described by the Lieb-Liniger model (LLM), which has been
exactly solved by the Bethe ansatz. We can compare the result of the IGA
method with the exact one in order to test the precision and validity of the
IGA method. In the future, we shall apply IGA to 2D or 3D Bose gas at finite
temperature, as in high dimension we can not apply the Bethe ansatz method
to obtain the exact solution, IGA or IDT is the only approach we can rely
on. In higher dimension, the quantum and thermal fluctuations are weaker
than in 1D, the result obtained by IGA or IDT should be better qualitatively
and quantitatively than that in 1D.

In this paper, we shall study LLM by using IGA, and focus our attention on
the excitation spectrum. We will follow \cite{Rosenstein} and present IGA
method by solving Dyson-Schwinger equations which are generated by
functional differentiation of the effective action.

We will show that only the Bogoliubov excitation spectrum (or Type I
excitation) can be obtained by IGA. By comparing with the results of the
Bogoliubov approximation and Type I excitation based on the exact solution
\cite{key-19,key-29}, we find that the spectrum obtained in this way is good
improvement to the spectrum in the Bogoliubov approximation. In order to
obtain Type II excitation, we speculate that we shall use more general $\Phi
$ derivable theory beyond IGA (we will leave it as our future work). If we
study high dimension Bosonic system, there will be no Type II excitation,
IGA will give more accurate results quantitatively.

The rest of the paper is organized as follows. In section II we review the
basic formulation of one particle irreducible (1PI) effective action theory
and the Dyson-Schwinger equations. We also present the 1D bosonic model and
the Dyson-Schwinger equations for 1D bosonic model in this section. In
section III we review the traditional approximations, such as Bogoliubov
approximation, HFB approximation and Popov approximation. In section IV, we
present improved Gaussian approximation and obtain an improved gapless
excitation spectrum. In section V we make a comparison with the exact
solution of the 1D bosonic model \cite{key-19,key-29}. Finally, we give a
summary and the conclusions. We put $\hbar =k_{B}=1$ throughout the paper
with $k_{B}$ the Boltzmann constant.

\section{The Dyson-Schwinger equations for 1D bosonic model}

In this section we shall present the general formulations and the model, and
set up all the notations and definitions. We shall start with the
thermodynamic partition function and set the temperature to zero in the end.
For a bosonic system, the grand canonical partition function takes the form
\cite{key-38}

\begin{equation}
\mathcal{Z}=\int \mathcal{D}[\psi ^{\ast }\psi ]e^{-S[\psi ^{\ast },\psi ]}
\end{equation}%
with the classical action $S[\psi ^{\ast },\psi ]$ given by

\begin{equation}
\int_{0}^{\beta }d\tau \int d^{D}\mathbf{x}\left( \psi ^{\ast }\partial
_{\tau }\psi -\mu \psi ^{\ast }\psi +\mathcal{H}\left[ \psi ^{\ast },\psi %
\right] \right)  \label{eq:action}
\end{equation}%
where $\beta =\frac{1}{k_{B}T}$ , $\mu $ is the chemical potential and $%
\mathcal{H}\left[ \psi ^{\ast },\psi \right] $ is the Hamiltonian density, $%
D $ is the dimension of position space (the formulation is valid for
arbitary $D$, however in this paper, we will only carry out calculations for
1D). In order to obtain the correlation functions of field operators, a
generating functional is defined by coupling fields to an external source,

\begin{equation}
\mathcal{Z}[J^{\ast },J]=\int \mathcal{D}[\psi ^{\ast },\psi ]e^{-\left(
S[\psi ^{\ast },\psi ]+J^{\ast }\psi +J\psi ^{\ast }\right) },
\end{equation}%
where $J^{\ast }\psi $ is a shorthand for $\int_{0}^{\beta }d\tau \int d^{D}%
\mathbf{x}J^{\ast }(\mathbf{x},\tau )\psi (\mathbf{x},\tau )$ and similarly
for $J\psi ^{\ast }$. The connected generating functional is defined as
\begin{equation}
W[J^{\ast },J]=-\ln \mathcal{Z}[J^{\ast },J].
\end{equation}%
The one-point expectation value of the field operators can be obtained by
the derivatives of the generating functional with respect to the external
source,%
\begin{eqnarray}
\varphi (\mathbf{x},\tau ) &=&\frac{\delta W[J^{\ast },J]}{\delta J^{\ast }(%
\mathbf{x},\tau )}  \notag \\
\varphi ^{\ast }(\mathbf{x},\tau ) &=&\frac{\delta W[J^{\ast },J]}{\delta J(%
\mathbf{x},\tau )}  \label{eq:WJ2}
\end{eqnarray}%
where $\varphi (\mathbf{x},\tau )=\left\langle \psi (\mathbf{x},\tau
)\right\rangle $, $\varphi ^{\ast }(\mathbf{x},\tau )=\left\langle \psi
^{\ast }(\mathbf{x},\tau )\right\rangle $ with
\begin{equation}
\left\langle \cdots \right\rangle \equiv \frac{1}{\mathcal{Z}[J^{\ast },J]}%
\int \mathcal{D}[\psi ^{\ast },\psi ]\cdots e^{-\left( S[\psi ^{\ast },\psi
]+J^{\ast }\psi +J\psi ^{\ast }\right) }.
\end{equation}%
Successive derivatives generate multi-point correlation functions, for
instance,
\begin{equation}
\frac{\delta ^{2}W}{\delta J(x)\delta J^{\ast }(y)}=-\left\langle \psi
^{\ast }(x)\psi (y)\right\rangle _{c}
\end{equation}%
where $x\equiv (\mathbf{x},\tau )$, $y\equiv (\mathbf{y},\tau ^{\prime })$
and the connected Green's function $\left\langle \psi ^{\ast }(x)\psi
(y)\right\rangle _{c}=$ $\left\langle \psi ^{\ast }(x)\psi (y)\right\rangle
-\left\langle \psi ^{\ast }(x)\right\rangle \left\langle \psi
(y)\right\rangle $. For notation compactness, we define
\begin{eqnarray}
(J,\,J^{\ast }) &\equiv &(J_{1},\,J_{2}),(\psi ^{\ast },\psi )\equiv (\psi
_{1},\psi _{2}),(\varphi ^{\ast },\varphi )\equiv (\varphi _{1},\varphi
_{2}),  \notag \\
G_{mn}(x,y) &\equiv &\left\langle \psi _{m}(x)\psi _{n}(y)\right\rangle _{c},
\label{eq:Gmn}
\end{eqnarray}%
where $m=1,2$, $n=1,2$. $G_{mn}(x,y)$ is related to $W[J^{\ast },J]$ by the
following equation,
\begin{equation}
G_{mn}(x,y)=-\frac{\delta ^{2}W}{\delta J_{m}(x)\delta J_{n}(y)}.
\end{equation}

The 1PI effective action is defined by the Legendre transformation,
\begin{equation}
\Gamma \lbrack \varphi ^{\ast },\varphi ]=W[J^{\ast },J]-J^{\ast }\varphi
-J\varphi ^{\ast },
\end{equation}%
which is a functional of the field expectation $\varphi ^{\ast }$and $%
\varphi $. In analogy with Eq.\eqref{eq:WJ2} the external source can be
obtained by the derivatives of the effective action with respect to the
one-point expectation of the field operators,

\begin{eqnarray}
\frac{\delta \Gamma \lbrack \varphi ^{\ast },\varphi ]}{\delta \varphi (%
\mathbf{x},\tau )} &=&-J^{\ast }(\mathbf{x},\tau ),  \notag \\
\frac{\delta \Gamma \lbrack \varphi ^{\ast },\varphi ]}{\delta \varphi
^{\ast }(\mathbf{x},\tau )} &=&-J(\mathbf{x},\tau ).  \label{eq:Gamma-eq2}
\end{eqnarray}%
The effective action is the generating functional for vertex functions.
Using the chain rule to calculate $\frac{\delta \varphi _{m}(x)}{\delta
\varphi _{n}(y)}$, we have
\begin{eqnarray}
\frac{\delta \varphi _{m}(x)}{\delta \varphi _{n}(y)} &=&\sum_{i}\int dz%
\frac{\delta \varphi _{m}(x)}{\delta J_{i}(z)}\frac{\delta J_{i}(z)}{\delta
\varphi _{n}(y)}  \notag \\
&=&-\sum_{i}\int dz\frac{\delta ^{2}W}{\delta J_{m}(x)\delta J_{i}(z)}\frac{%
\delta ^{2}\Gamma }{\delta \varphi _{i}(z)\delta \varphi _{n}(y)}.
\label{eq:chain}
\end{eqnarray}%
On the other hand,%
\begin{equation}
\frac{\delta \varphi _{m}(x)}{\delta \varphi _{n}(y)}=\delta _{mn}\delta
(x-y).  \label{eq:identity}
\end{equation}%
Thus by combining Eqs.\eqref{eq:chain}\eqref{eq:identity} one obtains
\begin{equation}
\sum_{i}\int dzG_{mi}(x,z)\Gamma _{in}(z,y)=\delta _{mn}\delta (x-y)
\label{eq:G-Gamma-I}
\end{equation}%
where $\Gamma _{mn}(x,y)\equiv \frac{\delta ^{2}\Gamma \lbrack \varphi
_{1},\varphi _{2}]}{\delta \varphi _{m}(x)\delta \varphi _{n}(y)}$ and $%
G_{mn}(x,y)$ is defined in Eq.\eqref{eq:Gmn}. The 1PI effective action $%
\Gamma \lbrack \varphi ^{\ast },\varphi ]$ can be approximately obtained by
loop expansion \cite{key-39}.

Dyson-Schwinger equations can be obtained by using the following identity,
\begin{equation}
\int \mathcal{D}[\psi ^{\ast },\psi ]\frac{\delta }{\delta \psi ^{\ast }(x)}%
e^{-\left( S[\psi ^{\ast },\psi ]+J^{\ast }\psi +J\psi ^{\ast }\right) }=0,
\end{equation}%
which leads to
\begin{equation}
\left\langle \frac{\delta S[\psi ^{\ast },\psi ]}{\delta \psi ^{\ast }(x)}%
\right\rangle +J(x)=0.  \label{eq:first}
\end{equation}
Derivatives of Eq.\eqref{eq:first} with respect to the average field $%
\varphi _{m}(x)$ shall produce a series of Dyson-Schwinger equations, such
as
\begin{equation}
\frac{\delta }{\delta \varphi (y)}\left\langle \frac{\delta S[\psi ^{\ast
},\psi ]}{\delta \psi ^{\ast }(x)}\right\rangle +\frac{\delta }{\delta
\varphi (y)}J(x)=0.  \label{eq:second}
\end{equation}%
Successive functional derivatives with respect to $\varphi \left( z\right) $
yield higher order Dyson-Schwinger equations, which involve the correlation
functions of more field operators. Therefore, the infinite Dyson-Schwinger
equations must be truncated to form a set of closed equations in order to
carry out any calculations. Let us term Eq.\eqref{eq:first} as the first
Dyson-Schwinger equation and Eq.\eqref{eq:second} as the second
Dyson-Schwinger equation.

We apply the Dyson-Schwinger formalism to a system of one-dimensional
bosonic gas interacting via a repulsive contact potential, described by the
Lieb-Liniger Hamiltonian%
\begin{equation}
H=-\sum_{i=1}^{N}\left( \partial ^{2}/\partial x_{i}^{2}\right)
+g\sum_{i<j}^{N}\delta (x_{i}-x_{j}),  \label{eq:Hamiltonian}
\end{equation}%
where the mass of the particle has been set to $2m=1$ and $g$ is the contact
interaction strength, which is related to the 1D scattering length
experimentally. The second quantization form reads
\begin{equation}
\hat{H}=\int d^{D}\mathbf{x}\left( \psi ^{\dagger }(\mathbf{x})(-\nabla
^{2})\psi (\mathbf{x})+\frac{1}{2}g\psi ^{\dagger }(\mathbf{x})\psi
^{\dagger }(\mathbf{x})\psi (\mathbf{x})\psi (\mathbf{x})\right) ,
\end{equation}%
where we have used the notation for a general position space dimension $D$
and bear in mind that we will study the 1D case of $D=1$ in the end.

In path-integral formalism, the grand canonical partition function takes the
form

\begin{equation}
\mathcal{Z}=\int \mathcal{D}[\psi ^{\ast }\psi ]e^{-S[\psi ^{\ast },\psi ]}
\end{equation}%
with the classical action $S[\psi ^{\ast },\psi ]$ given by

\begin{equation}
\int_{0}^{\beta }d\tau \int d^{D}\mathbf{x}\left( \psi ^{\ast }\left(
\partial _{\tau }-\mu -\nabla ^{2}\right) \psi +\frac{1}{2}g\psi ^{\ast
}\psi ^{\ast }\psi \psi \right)  \label{eq:action-1}
\end{equation}%
where $\psi \equiv \psi (\mathbf{x},\tau )$, $\beta =\frac{1}{k_{B}T}$ and $%
\mu $ is the chemical potential. By variable rescaling
\begin{eqnarray}
\psi &=&\sqrt{g}\psi ^{\prime },\tau =g^{-2}\tau ^{\prime },  \notag \\
\mathbf{x} &=&g^{-1}\mathbf{x^{\prime }},\mu =g^{2}\mu ^{\prime },
\label{eq:rescale}
\end{eqnarray}%
the action can be recast as a simple form dependent only on one parameter $%
\mu ^{\prime }$,%
\begin{equation}
\int_{0}^{\beta ^{\prime }}d\tau ^{\prime }\int d^{D}\mathbf{x}^{\prime
}\left( \psi ^{\prime \ast }\left( \partial _{\tau ^{\prime }}-\nabla _{%
\mathbf{x}^{\prime }}^{2}-\mu ^{\prime }\right) \psi ^{\prime }+\frac{1}{2}%
\psi ^{\prime \ast }\psi ^{\prime \ast }\psi ^{\prime }\psi ^{\prime
}\right) .  \label{eq:action-prime}
\end{equation}%
In the following discussions, we will omit the primes for simplicity,
\begin{equation}
S\left[ \psi ^{\ast },\psi \right] =\int_{0}^{\beta }d\tau \int d^{D}\mathbf{%
x}\left( \psi ^{\ast }\left( \partial _{\tau }-\nabla ^{2}-\mu \right) \psi +%
\frac{1}{2}\psi ^{\ast }\psi ^{\ast }\psi \psi \right) .
\label{eq:rescaled-action}
\end{equation}%
Starting with the rescaled action in Eq.\eqref{eq:rescaled-action}, we
define the generating functional

\begin{equation}
\mathcal{Z}[J^{\ast },J]=\int \mathcal{D}[\psi ^{\ast },\psi ]e^{-\left(
S[\psi ^{\ast },\psi ]+J^{\ast }\psi +J\psi ^{\ast }\right) }.
\end{equation}
The first Dyson-Schwinger equations take the form

\begin{eqnarray}
\left( \partial _{\tau }-\nabla ^{2}-\mu \right) \varphi _{2}+\left\langle
\psi _{1}\psi _{2}\psi _{2}\right\rangle +J_{1} &=&0,  \notag \\
\left( -\partial _{\tau }-\nabla ^{2}-\mu \right) \varphi _{1}+\left\langle
\psi _{1}\psi _{1}\psi _{2}\right\rangle +J_{2} &=&0,  \label{eq:motion2}
\end{eqnarray}%
where implicitly all the arguments are $x\equiv (\mathbf{x},\tau )$.
By Wick theorem we know
\begin{equation}
\left\langle \psi _{1}\psi _{2}\psi _{2}\right\rangle =\left\langle
\psi _{1}\psi _{2}\psi _{2}\right\rangle _{c}+2\varphi
_{2}\left\langle \psi _{1}\psi _{2}\right\rangle _{c}+\varphi
_{1}\left\langle \psi _{2}\psi _{2}\right\rangle _{c}+\varphi
_{1}\varphi _{2}^{2}\, , \label{eq:Wick}
\end{equation}%
where $\left\langle \cdots \right\rangle _{c}$ means connected
correlation functions. Substituting Eq.\eqref{eq:Wick} into
Eq.\eqref{eq:motion2} yields

\begin{eqnarray}
\left( \partial _{\tau }-\nabla ^{2}-\mu \right) \varphi _{2}+\varphi
_{1}\varphi _{2}^{2}+\varphi _{1}G_{22}+2\varphi _{2}G_{12}+\left\langle
\psi _{1}\psi _{2}\psi _{2}\right\rangle _{c}+J_{1} &=&0,  \notag \\
\left( -\partial _{\tau }-\nabla ^{2}-\mu \right) \varphi _{1}+\varphi
_{1}^{2}\varphi _{2}+\varphi _{2}G_{11}+2\varphi _{1}G_{12}+\left\langle
\psi _{2}\psi _{1}\psi _{1}\right\rangle _{c}+J_{2} &=&0,  \label{eq:first-2}
\end{eqnarray}%
where all the default arguments are $x\equiv (\mathbf{x},\tau )$ and $%
G_{11}=G_{11}(x,x)$ , $G_{22}=G_{22}(x,x)$, $G_{12}=\left\langle \psi
_{1}(x)\psi _{2}(x)\right\rangle _{c}$. $G_{ij}=G_{ij}(x,x)$ is a constant
for a translational symmetric system which is the case in this paper.
Further differentiations of Eq.\eqref{eq:first-2} with respect to $\varphi
_{1}(y)$ and $\varphi _{2}(y)$ result in the second Dyson-Schwinger
equations,
\begin{gather}
\Gamma _{11}(x,y)=\left( \varphi _{2}^{2}+G_{22}\right) \delta (x-y)  \notag
\\
+\varphi _{1}\Lambda _{221}(x,y)+2\varphi _{2}\Lambda _{121}(x,y)+\frac{%
\delta }{\delta \varphi _{1}(y)}\left\langle \psi _{1}\psi _{2}\psi
_{2}\right\rangle _{c},  \notag \\
\Gamma _{22}(x,y)=\left( \varphi _{1}^{2}+G_{11}\right) \delta (x-y)  \notag
\\
+\varphi _{2}\Lambda _{112}(x,y)+2\varphi _{1}\Lambda _{122}(x,y)+\frac{%
\delta }{\delta \varphi _{2}(y)}\left\langle \psi _{2}\psi _{1}\psi
_{1}\right\rangle _{c},  \notag \\
\Gamma _{12}(x,y)=\left( \partial _{\tau }-\nabla _{\mathbf{x}}^{2}-\mu
+2\varphi _{1}\varphi _{2}+2G_{12}\right) \delta (x-y)  \notag \\
+\varphi _{1}\Lambda _{222}(x,y)+2\varphi _{2}\Lambda _{122}(x,y)+\frac{%
\delta }{\delta \varphi _{2}(y)}\left\langle \psi _{1}\psi _{2}\psi
_{2}\right\rangle _{c},  \notag \\
\Gamma _{21}(x,y)=\left( -\partial _{\tau }-\nabla _{\mathbf{x}}^{2}-\mu
+2\varphi _{1}\varphi _{2}+2G_{12}\right) \delta (x-y)  \notag \\
+\varphi _{2}\Lambda _{111}(x,y)+2\varphi _{1}\Lambda _{121}(x,y)+\frac{%
\delta }{\delta \varphi _{1}(y)}\left\langle \psi _{2}\psi _{1}\psi
_{1}\right\rangle _{c},  \label{eq:second-21}
\end{gather}%
where $x\equiv (\mathbf{x},\tau )$, $y\equiv (\mathbf{y},\tau ^{\prime })$
and $\Lambda _{mnl}(x,y)\equiv \frac{\delta G_{mn}(x,x)}{\delta \varphi
_{l}(y)}$ with $m\,(n,\,l)=1,\,2$. Since what we consider is a homogeneous
gas, we can set
\begin{equation}
\varphi _{1}(\mathbf{x},\tau )=\varphi _{2}(\mathbf{x},\tau )\equiv \upsilon
,  \label{eq:homo-condition}
\end{equation}%
where $\upsilon $ is a real constant number. Further, we define the Fourier
transformations%
\begin{eqnarray}
\delta (x-y) &=&\int \frac{d\omega }{2\pi }\int \frac{d^{D}\mathbf{k}}{(2\pi
)^{D}}e^{i\mathbf{k}\cdot (\mathbf{x}-\mathbf{y})-i\omega (\tau -\tau
^{\prime })},  \notag \\
\Lambda _{mnl}(x,y) &=&\int \frac{d\omega }{2\pi }\int \frac{d^{D}\mathbf{k}%
}{(2\pi )^{D}}\Lambda _{mnl}(k)e^{i\mathbf{k}\cdot (\mathbf{x}-\mathbf{y}%
)-i\omega (\tau -\tau ^{\prime })},  \notag \\
G_{mn}(x,y) &=&\int \frac{d\omega }{2\pi }\int \frac{d^{D}\mathbf{k}}{(2\pi
)^{D}}G_{mn}(k)e^{i\mathbf{k}\cdot (\mathbf{x}-\mathbf{y})-i\omega (\tau
-\tau ^{\prime })},  \notag \\
\Gamma _{mn}(x,y) &=&\int \frac{d\omega }{2\pi }\int \frac{d^{D}\mathbf{k}}{%
(2\pi )^{D}}\Gamma _{mn}(k)e^{i\mathbf{k}\cdot (\mathbf{x}-\mathbf{y}%
)-i\omega (\tau -\tau ^{\prime })},  \label{eq:FT}
\end{eqnarray}%
where $k\equiv (\mathbf{k},\omega )$ and $\omega $ denotes the Matsubara
frequency in the zero temperature limit. In the frequency space, Eq.%
\eqref{eq:G-Gamma-I} is recast as

\begin{equation}
\sum_{m=1,2}G_{im}(k)\Gamma _{mj}(k)=\delta _{ij}.  \label{eq:G-Gamma-k}
\end{equation}

The first and second Dyson-Schwinger equations are not closed equations.
They are impossible to solve unless truncations are performed.

\section{The traditional approximations}

The traditional approximations, such as Bogoliubov approximation, HFB
approximation and Popov approximation, have been exhaustively discussed in
the literature. In order to clarify the interrelations of the various
familiar schemes and the IGA scheme we shall present later, in this section
we formulate those approximations by truncating the first and second
Dyson-Schwinger equations.

\paragraph{Bogoliubov approximation:}

Ignoring any correlations, only the first Dyson-Schwinger equations Eq.%
\eqref{eq:first-2} are retained:
\begin{eqnarray}
\left( \partial _{\tau }-\nabla ^{2}-\mu \right) \varphi _{2}+\varphi
_{1}\varphi _{2}^{2}+J_{1} &=&0,  \notag \\
\left( -\partial _{\tau }-\nabla ^{2}-\mu \right) \varphi _{1}+\varphi
_{1}^{2}\varphi _{2}+J_{2} &=&0,  \label{eq:B2}
\end{eqnarray}%
and the two-point vertex functions are defined by $\Gamma _{ij}(x,y)=-\frac{%
\delta J_{i}(x)}{\delta \varphi _{j}(y)}|_{J_{i}(x)=0}$ where $J_{i}(x)$, $%
\varphi _{j}(y)$ are related by Eq.\eqref{eq:B2},
\begin{eqnarray}
\Gamma _{11}(x,y) &=&\varphi _{2}^{2}\delta (x-y),  \notag \\
\Gamma _{22}(x,y) &=&\varphi _{1}^{2}\delta (x-y),  \notag \\
\Gamma _{12}(x,y) &=&\left( \partial _{\tau }-\nabla _{\mathbf{x}}^{2}-\mu
+2\varphi _{1}\varphi _{2}\right) \delta (x-y),  \notag \\
\Gamma _{21}(x,y) &=&\left( -\partial _{\tau }-\nabla _{\mathbf{x}}^{2}-\mu
+2\varphi _{1}\varphi _{2}\right) \delta (x-y).  \label{eq:B21}
\end{eqnarray}%
By using the homogeneous and static condition in Eq.\eqref{eq:homo-condition}
and applying the Fourier transformation in Eq.\eqref{eq:FT}, we rewrite Eq.%
\eqref{eq:B2} when $J_{i}(x)=0$ as%
\begin{equation}
\upsilon ^{2}=\mu  \label{eq:B1u}
\end{equation}%
and Eq.\eqref{eq:B21} becomes when $J_{i}(x)=0$,

\begin{eqnarray}
\Gamma _{11}(k) &=&\upsilon ^{2},\Gamma _{22}(k)=\upsilon ^{2},  \notag \\
\Gamma _{12}(k) &=&-i\omega +\mathbf{k}^{2}+\upsilon ^{2},  \notag \\
\Gamma _{21}(k) &=&i\omega +\mathbf{k}^{2}+\upsilon ^{2}.
\end{eqnarray}

With the help of Eq.\eqref{eq:G-Gamma-k} we obtain the Green's functions in
Bogoliubov approximation,
\begin{gather}
\left(
\begin{array}{cc}
G_{11}(k) & G_{12}(k) \\
G_{21}(k) & G_{22}(k)%
\end{array}%
\right) =\frac{1}{(i\omega )^{2}-\mathbf{k}^{2}\left( \mathbf{k}%
^{2}+2\upsilon ^{2}\right) }  \notag \\
\times \left(
\begin{array}{cc}
\upsilon ^{2} & i\omega -\left( \mathbf{k}^{2}+\upsilon ^{2}\right) \\
-i\omega -\left( \mathbf{k}^{2}+\upsilon ^{2}\right) & \upsilon ^{2}%
\end{array}%
\right) .  \label{eq:G-Bog}
\end{gather}%
The Bogoliubov spectrum is given by the pole of the determinant of Matrix Eq.%
\eqref{eq:G-Bog}

\begin{equation}
\varepsilon _{\text{Bog}}(k)=k\sqrt{k^{2}+2\upsilon ^{2}}.
\end{equation}%
In this approximation, the particle density $n$ is equal to $\upsilon ^{2}$.

\paragraph{HFB approximation:}

If two-point correlation functions are kept, ignoring three or higher point
correlation functions, the first Dyson-Schwinger equations Eq.%
\eqref{eq:first-2} become

\begin{eqnarray}
\left( \partial _{\tau }-\nabla ^{2}-\mu \right) \varphi _{2}+\varphi
_{1}\varphi _{2}^{2}+\varphi _{1}G_{22}+2\varphi _{2}G_{12} &=&0,  \notag \\
\left( -\partial _{\tau }-\nabla ^{2}-\mu \right) \varphi _{1}+\varphi
_{1}^{2}\varphi _{2}+\varphi _{2}G_{11}+2\varphi _{1}G_{12} &=&0,
\label{eq:HFB-shift}
\end{eqnarray}%
and the second Dyson-Schwinger equations Eq.\eqref{eq:second-21} become
\begin{eqnarray}
\Gamma _{11}(x,y) &=&\left( \varphi _{2}^{2}+G_{22}\right) \delta (x-y),
\notag \\
\Gamma _{22}(x,y) &=&\left( \varphi _{1}^{2}+G_{11}\right) \delta (x-y),
\notag \\
\Gamma _{12}(x,y) &=&\left( \partial _{\tau }-\nabla _{\mathbf{x}}^{2}-\mu
+2\varphi _{1}\varphi _{2}+2G_{12}\right) \delta (x-y),  \notag \\
\Gamma _{21}(x,y) &=&\left( -\partial _{\tau }-\nabla _{\mathbf{x}}^{2}-\mu
+2\varphi _{1}\varphi _{2}+2G_{12}\right) \delta (x-y).  \label{eq:HFB-gap}
\end{eqnarray}

By using the homogeneous and static condition in Eq.\eqref{eq:homo-condition}
and applying the Fourier transformation in Eq.\eqref{eq:FT}, we rewrite Eq.%
\eqref{eq:HFB-shift} as%
\begin{equation}
0=-\mu +\upsilon ^{2}+G_{11}+2G_{12},G_{11}=G_{22},
\end{equation}%
and Eq.\eqref{eq:HFB-gap} as

\begin{eqnarray}
\Gamma _{11}(k) &=&\upsilon ^{2}+G_{11},  \notag \\
\Gamma _{22}(k) &=&\upsilon ^{2}+G_{11},  \notag \\
\Gamma _{12}(k) &=&-i\omega +\mathbf{k}^{2}+\upsilon ^{2}-G_{11},  \notag \\
\Gamma _{21}(k) &=&i\omega +\mathbf{k}^{2}+\upsilon ^{2}-G_{11}.
\end{eqnarray}%
With the help of Eq.\eqref{eq:G-Gamma-k} we obtain the two-point Green's
functions in HFB approximation,%
\begin{gather}
\left(
\begin{array}{cc}
G_{11}(k) & G_{12}(k) \\
G_{21}(k) & G_{22}(k)%
\end{array}%
\right) =\frac{1}{(i\omega )^{2}-\left( \mathbf{k}^{2}+2\upsilon ^{2}\right)
\left( \mathbf{k}^{2}-2G_{11}\right) }  \notag \\
\times \left(
\begin{array}{cc}
\upsilon ^{2}+G_{11} & i\omega -\left( \mathbf{k}^{2}+\upsilon
^{2}-G_{11}\right) \\
-i\omega -\left( \mathbf{k}^{2}+\upsilon ^{2}-G_{11}\right) & \upsilon
^{2}+G_{11}%
\end{array}%
\right) .  \label{eq:tr-G}
\end{gather}%
Then the HFB spectrum is given by

\begin{equation}
\varepsilon _{\text{HFB}}(\mathbf{k})=\sqrt{\left( \mathbf{k}^{2}+2\upsilon
^{2}\right) \left( \mathbf{k}^{2}-2G_{11}\right) }.  \label{eq:eHFB}
\end{equation}%
The variable $G_{11}$ can be determined in a self-consistent way. By the
definitions of $G_{11}$and $G_{12}$, there are
\begin{equation}
G_{11}=-\frac{1}{4\pi }\left( \upsilon ^{2}+G_{11}\right) \int_{-\infty
}^{\infty }dk\frac{1}{\sqrt{\left( k^{2}+2\upsilon ^{2}\right) \left(
k^{2}-2G_{11}\right) }},  \label{eq:G11}
\end{equation}%
and

\begin{equation}
G_{12}=\frac{1}{4\pi }\int_{-\infty }^{\infty }dk\left( \frac{\left(
k^{2}+\upsilon ^{2}-G_{11}\right) }{\sqrt{\left( k^{2}+2\upsilon ^{2}\right)
\left( k^{2}-2G_{11}\right) }}-1\right) .  \label{eq:G12}
\end{equation}

In HFB approximation, the particle number density is
\begin{equation}
n=\upsilon ^{2}+G_{12}  \label{eq:number}
\end{equation}

\paragraph{Popov approximation:}

Popov approximation is well-known for its gapless excitation spectrum. It
differs from the HFB approximation in neglecting the {}\textquotedblleft
anomalous\textquotedblright\ two-point correlations $G_{11}$and $G_{22}$, so
that the Dyson-Schwinger equations take the form
\begin{equation}
-\mu +\upsilon ^{2}+2G_{12}=0
\end{equation}%
and%
\begin{eqnarray}
\Gamma _{11}(k) &=&\upsilon ^{2},\Gamma _{22}(k)=\upsilon ^{2},  \notag \\
\Gamma _{12}(k) &=&-i\omega +\mathbf{k}^{2}+\upsilon ^{2},\Gamma
_{21}(k)=i\omega +\mathbf{k}^{2}+\upsilon ^{2}.
\end{eqnarray}%
In terms of the variable $\upsilon ^{2}$, the two-point Green's functions
have the similar form as those in the Bogoliubov approximation,
\begin{gather}
\left(
\begin{array}{cc}
G_{11}(k) & G_{12}(k) \\
G_{21}(k) & G_{22}(k)%
\end{array}%
\right) =\frac{1}{(i\omega )^{2}-\mathbf{k}^{2}\left( \mathbf{k}%
^{2}+2\upsilon ^{2}\right) }  \notag \\
\times \left(
\begin{array}{cc}
\upsilon ^{2} & i\omega -\left( \mathbf{k}^{2}+\upsilon ^{2}\right) \\
-i\omega -\left( \mathbf{k}^{2}+\upsilon ^{2}\right) & \upsilon ^{2}%
\end{array}%
\right) .  \label{eq:Popov}
\end{gather}%
and also the excitation spectrum%
\begin{equation}
\varepsilon _{\text{Popov}}(k)=k\sqrt{k^{2}+2\upsilon ^{2}}.
\end{equation}%
By the definition of $G_{12}$, there is
\begin{equation}
G_{12}=\int \frac{d\omega }{2\pi }\int \frac{d^{D}\mathbf{k}}{(2\pi )^{D}}%
\frac{i\omega -\left( \mathbf{k}^{2}+\upsilon ^{2}\right) }{(i\omega )^{2}-%
\mathbf{k}^{2}\left( \mathbf{k}^{2}+2\upsilon ^{2}\right) }.
\end{equation}%
The particle number density is given by $n=\upsilon ^{2}+G_{12}$.
However, in 1D, the above equation leads to
\begin{equation}
n=\upsilon ^{2}+\frac{1}{4\pi }\int_{-\infty }^{\infty }dk\left( -1+\frac{1}{%
k}\sqrt{k^{2}+2\upsilon ^{2}}-\frac{\upsilon ^{2}}{k\sqrt{k^{2}+2\upsilon
^{2}}}\right) ,  \label{eq:n12-popov-1}
\end{equation}%
which is infrared divergent. So the Popov approximation is inapplicable here.

The reason for Popov theory to break down in 1D is that phase fluctuations
are not considered properly. Ref.\cite{Stoof} gave a detailed discussion of
this problem and proposed the modified Popov theory, in which the
inappropriately incorporated phase fluctuations are subtracted and thus the
infrared divergence is removed. The particle number density from the
modified Popov theory shall be given by
\begin{equation}
n=\upsilon ^{2}+\frac{1}{4\pi }\int_{-\infty }^{\infty }dk\left( -1+\frac{k}{%
\sqrt{k^{2}+2\upsilon ^{2}}}\right) ,  \label{eq:M-Popov-n}
\end{equation}%
which is free of divergences.

\section{Improved Gaussian approximation}

In this section we shall present another strategy, IGA (improved Gaussian
approximation) which takes account of quantum fluctuations more precisely
(adding some Feymann diagrams to preserve symmetry requirement) and retains
the gapless Goldstone mode.

\begin{figure}[t]
\centering \rotatebox{0}{\includegraphics[scale=0.4]{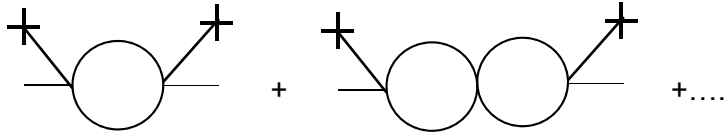}}
\caption{Feynman diagrams for the corrections to the two-point vertex
function obtained by HFB approximation.}
\label{fig1}
\end{figure}

By preserving up to two-point correlation functions in the first
Dyson-Schwinger equations, however we will keep source terms here for a
while in order to define the Green's function in IGA scheme.
\begin{eqnarray}
\left( \partial _{\tau }-\nabla ^{2}-\mu \right) \varphi _{2}+\varphi
_{1}\varphi _{2}^{2}+\varphi _{1}G_{22}^{tr}+2\varphi _{2}G_{12}^{tr}+J_{1}
&=&0,  \notag \\
\left( -\partial _{\tau }-\nabla ^{2}-\mu \right) \varphi _{1}+\varphi
_{1}^{2}\varphi _{2}+\varphi _{2}G_{11}^{tr}+2\varphi _{1}G_{12}^{tr}+J_{2}
&=&0,  \label{eq:sol-shift}
\end{eqnarray}%
and

\begin{eqnarray}
\Gamma _{11}^{tr}(x,y) &=&\left( \varphi _{2}^{2}+G_{22}^{tr}\right) \delta
(x-y)  \notag \\
\Gamma _{22}^{tr}(x,y) &=&\left( \varphi _{1}^{2}+G_{11}^{tr}\right) \delta
(x-y),  \notag \\
\Gamma _{12}^{tr}(x,y) &=&\left( \partial _{\tau }-\nabla _{\mathbf{x}%
}^{2}-\mu +2\varphi _{1}\varphi _{2}+2G_{12}^{tr}\right) \delta (x-y)  \notag
\\
\Gamma _{21}^{tr}(x,y) &=&\left( -\partial _{\tau }-\nabla _{\mathbf{x}%
}^{2}-\mu +2\varphi _{1}\varphi _{2}+2G_{12}^{tr}\right) \delta (x-y),
\label{eq:gapeq}
\end{eqnarray}

where $tr$ is the abbreviation of "truncation". \ We will define \bigskip
\begin{equation}
\Gamma _{ij}(x,y)=-\frac{\delta J_{i}(x)}{\delta \varphi _{j}(y)}%
|_{J_{i}(x)=0}
\end{equation}%
where the relations between $J_{i}\left( x\right) $ and $\varphi _{j}(y)$
are given by Eqs.(\ref{eq:sol-shift},\ref{eq:gapeq}).

\begin{eqnarray}
\Gamma _{11}(x,y) &=&\left( \varphi _{2}^{2}+G_{22}^{tr}\right) \delta
(x-y)+\varphi _{1}\Lambda _{221}^{tr}(x,y)+2\varphi _{2}\Lambda
_{121}^{tr}(x,y),  \notag \\
\Gamma _{22}(x,y) &=&\left( \varphi _{1}^{2}+G_{11}^{tr}\right) \delta
(x-y)+\varphi _{2}\Lambda _{112}^{tr}(x,y)+2\varphi _{1}\Lambda
_{122}^{tr}(x,y),  \notag \\
\Gamma _{12}(x,y) &=&\left( \partial _{\tau }-\nabla _{\mathbf{x}}^{2}-\mu
+2\varphi _{1}\varphi _{2}+2G_{12}^{tr}\right) \delta (x-y)+\varphi
_{1}\Lambda _{222}^{tr}(x,y)+2\varphi _{2}\Lambda _{122}^{tr}(x,y),  \notag
\\
\Gamma _{21}(x,y) &=&\left( -\partial _{\tau }-\nabla _{\mathbf{x}}^{2}-\mu
+2\varphi _{1}\varphi _{2}+2G_{12}^{tr}\right) \delta (x-y)  \notag \\
&&+\varphi _{2}\Lambda _{111}^{tr}(x,y)+2\varphi _{1}\Lambda
_{121}^{tr}(x,y),  \label{eq:sol-gap}
\end{eqnarray}%
where%
\begin{equation}
\Lambda _{mnl}^{tr}(x,y)\equiv \frac{\delta G_{mn}^{tr}(x,x)}{\delta \varphi
_{l}(y)},  \label{Lambda-tr}
\end{equation}%
and in the end we shall take $J_{i}(x)=0$. From $\Gamma _{ij}(x,y)$, we can
obtain the Green's function which is the inverse of $\Gamma _{ij}(x,y)$. The
result obtained is gapless \cite{key-8,key-32}. $\Gamma _{ij}^{tr}(x,y)$ is
obtained from the truncated Dyson-Schwinger equation ignoring three-point
Green's function. We comment that diagrammatically the corrections $\Delta \Gamma
_{ij}(x,y)=\Gamma _{ij}(x,y)-\Gamma _{ij}^{tr}(x,y)$ correspond to some
additional diagrams \cite{Rosenstein, Okopinska, Hendrik}, which are plotted
schematically in Fig. \ref{fig1}. In the Feynman rules  of Fig. \ref{fig1},   the point vertices are defined by the
interaction part of $S\left[ \varphi _{1}+\psi _{1},\varphi _{2}+\psi _{2}\right] $,
i.e., the part with three and four $\psi _{i}$ fields expanded around $\varphi _{i}$. The lines in Fig. \ref{fig1}
stand for the
truncated Green's function $G_{ij}^{tr}$. The cross  in Fig. \ref{fig1}represents  $\varphi _{i}$  (details
can be found in Ref.\cite{Hendrik}).
By using the homogeneous and static condition in Eq.\eqref{eq:homo-condition}
and applying the Fourier transformation in Eq.\eqref{eq:FT}, we rewrite Eq.%
\eqref{eq:sol-gap} as
\begin{eqnarray}
\Gamma _{11}(k) &=&\upsilon ^{2}+G_{11}^{tr}+\upsilon \Lambda
_{221}^{tr}(k)+2\upsilon \Lambda _{121}^{tr}(k),  \notag \\
\Gamma _{22}(k) &=&\upsilon ^{2}+G_{11}^{tr}+\upsilon \Lambda
_{112}^{tr}(k)+2\upsilon \Lambda _{122}^{tr}(k),  \notag \\
\Gamma _{12}(k) &=&-i\omega +\mathbf{k}^{2}+\upsilon
^{2}-G_{11}^{tr}+\upsilon \Lambda _{222}^{tr}(k)+2\upsilon \Lambda
_{122}^{tr}(k),  \notag \\
\Gamma _{21}(k) &=&i\omega +\mathbf{k}^{2}+\upsilon
^{2}-G_{11}^{tr}+\upsilon \Lambda _{111}^{tr}(k)+2\upsilon \Lambda
_{121}^{tr}(k).  \label{eq:sol-gap-k}
\end{eqnarray}%
$\upsilon ^{2}$ and $G_{ij}^{tr}$ in the above equation are obtained from
the HFB equations in the previous section as in the end we take $J_{i}(x)=0$.

We start to calculate $\Lambda _{mnl}^{tr}(x,y)$. First, we differentiate Eq.%
\eqref{eq:gapeq} with respect to $\varphi _{m}(z)$,

\begin{eqnarray}
\Gamma _{111}^{tr}(x,y,z) &=&\delta (x-y)\Lambda _{221}^{tr}(x,x,z),  \notag
\\
\Gamma _{221}^{tr}(x,y,z) &=&\delta (x-y)\left( \Lambda
_{111}^{tr}(x,x,z)+2\varphi _{1}\delta (x-z)\right) ,  \notag \\
\Gamma _{121}^{tr}(x,y,z) &=&\delta (x-y)\left( 2\Lambda
_{121}^{tr}(x,x,z)+2\varphi _{2}\delta (x-z)\right) ,  \notag \\
\Gamma _{211}^{tr}(x,y,z) &=&\delta (x-y)\left( 2\Lambda
_{121}^{tr}(x,x,z)+2\varphi _{2}\delta (x-z)\right) ,  \notag \\
\Gamma _{222}^{tr}(x,y,z) &=&\delta (x-y)\Lambda _{112}^{tr}(x,x,z),  \notag
\\
\Gamma _{112}^{tr}(x,y,z) &=&\delta (x-y)\left( \Lambda
_{222}^{tr}(x,x,z)+2\varphi _{2}\delta (x-z)\right) ,  \notag \\
\Gamma _{122}^{tr}(x,y,z) &=&\delta (x-y)\left( 2\Lambda
_{122}^{tr}(x,x,z)+2\varphi _{1}\delta (x-z)\right) ,  \notag \\
\Gamma _{212}^{tr}(x,y,z) &=&\delta (x-y)\left( 2\Lambda
_{122}^{tr}(x,x,z)+2\varphi _{1}\delta (x-z)\right) ,  \label{eq:correct}
\end{eqnarray}%
where $z\equiv (\mathbf{z},\tau ^{\prime \prime })$. From Eq.%
\eqref{eq:G-Gamma-I} we know
\begin{equation}
G_{mn}^{tr}(x,y)=\sum_{m^{\prime }}\sum_{n^{\prime }}\int dx^{\prime }\int
dy^{\prime }G_{mm^{\prime }}^{tr}(x,x^{\prime })\Gamma _{m^{\prime
}n^{\prime }}^{tr}(x^{\prime },y^{\prime })G_{n^{\prime }n}^{tr}(y^{\prime
},y).
\end{equation}%
The derivatives of the above equation with respect to $\varphi _{l}(z)$
result in
\begin{equation}
\Lambda _{mnl}^{tr}(x,y,z)=-\sum_{m^{\prime }}\sum_{n^{\prime }}\int
dx^{\prime }\int dy^{\prime }G_{mm^{\prime }}^{tr}(x,x^{\prime })\Gamma
_{m^{\prime }n^{\prime }l}^{tr}(x^{\prime },y^{\prime },z)G_{n^{\prime
}n}^{tr}(y^{\prime },y).  \label{eq:laggag}
\end{equation}%
One can now take $J_{i}(x)=0$. $G_{mm^{\prime }}^{tr}(x,x^{\prime })$ is
thus given by HFB approximation in the above equation. By substituting Eq.%
\eqref{eq:correct} into Eq.\eqref{eq:laggag} and setting $x=y$, one obtains
a set of closed equations for $\Lambda _{mnl}(x,x,z)$,

\begin{gather}
\Lambda _{mnl}(x,x,z)=-2\upsilon \left[ G_{ml}^{tr}(x,z)G_{\bar{l}%
n}^{tr}(z,x)+G_{m\bar{l}}^{tr}(x,z)G_{ln}^{tr}(z,x)+G_{m\bar{l}}^{tr}(x,z)G_{%
\bar{l}n}^{tr}(z,x)\right]  \notag \\
-2\int dx^{\prime }\left( G_{ml}^{tr}(x,x^{\prime })G_{\bar{l}%
n}^{tr}(x^{\prime },x)\Lambda _{\bar{l}ll}(x^{\prime },x^{\prime },z)+G_{m%
\bar{l}}^{tr}(x,x^{\prime })G_{ln}^{tr}(x^{\prime },x)\Lambda _{\bar{l}%
ll}(x^{\prime },x^{\prime },z)\right)  \notag \\
-\int dx^{\prime }\left( G_{ml}^{tr}(x,x^{\prime })G_{ln}^{tr}(x^{\prime
},x)\Lambda _{\bar{l}\bar{l}l}(x^{\prime },x^{\prime },z)+G_{m\bar{l}%
}^{tr}(x,x^{\prime })G_{\bar{l}n}^{tr}(x^{\prime },x)\Lambda
_{lll}(x^{\prime },x^{\prime },z)\right) ,  \label{eq:self-con}
\end{gather}%
where $\bar{l}$ is defined by $\delta _{l\bar{l}}=0$, which means $l=1$, $%
\bar{l}=2$ or $l=2$, $\bar{l}=1$. By applying the Fourier transformations in
Eq.\eqref{eq:FT} we rewrite Eq. \eqref{eq:self-con} as

\begin{eqnarray}
\Lambda _{mnl}^{tr}(k) &=&\Lambda _{lll}^{tr}(k)I_{m\bar{l},\bar{l}%
n}(k)+\Lambda _{\bar{l}\bar{l}l}^{tr}(k)I_{ml,ln}(k)+\Lambda _{\bar{l}%
ll}^{tr}(k)\left( 2I_{ml,\bar{l}n}(k)+2I_{m\bar{l},ln}(k)\right)  \notag \\
&&+2\upsilon \left( I_{ml,\bar{l}n}(k)+I_{m\bar{l},ln}(k)+I_{m\bar{l},\bar{l}%
n}(k)\right) ,  \label{eq:self-con-k}
\end{eqnarray}%
where
\begin{equation}
I_{mn,m^{\prime }n^{\prime }}(k)=-\int \frac{d\omega _{1}}{2\pi }\int \frac{%
d^{D}\mathbf{k}_{1}}{(2\pi )^{D}}G_{mn}^{tr}(k_{1}+k)G_{m^{\prime }n^{\prime
}}^{tr}(k_{1})  \label{eq:fish}
\end{equation}%
and the two-point functions $G_{mn}^{tr}(k)$ are those obtained from the HFB
equations. We can explicitly integrate $\omega _{1}$ in Eq.(\ref{eq:fish}),
for example,

\begin{gather}
I_{11,11}(k)=\frac{1}{4}\int \frac{d^{D}\mathbf{k}_{1}}{(2\pi )^{D}}\frac{%
-(\upsilon ^{2}+G_{11}^{tr})^{2}}{\sqrt{\left( (k+\mathbf{k}%
_{1})^{2}+2\upsilon ^{2}\right) \left( (k+\mathbf{k}_{1})^{2}-2G_{11}^{tr}%
\right) (\mathbf{k}_{1}^{2}+2\upsilon ^{2})(\mathbf{k}_{1}^{2}-2G_{11}^{tr})}%
}  \notag \\
\times (\frac{1}{i\omega +\sqrt{\left( (k+\mathbf{k}_{1})^{2}+2\upsilon
^{2}\right) \left( (k+\mathbf{k}_{1})^{2}-2G_{11}^{tr}\right) }+\sqrt{(%
\mathbf{k}_{1}^{2}+2\upsilon ^{2})(\mathbf{k}_{1}^{2}-2G_{11}^{tr})}}  \notag
\\
-\frac{1}{i\omega -\sqrt{\left( (k+\mathbf{k}_{1})^{2}+2\upsilon ^{2}\right)
\left( (k+\mathbf{k}_{1})^{2}-2G_{11}^{tr}\right) }-\sqrt{(\mathbf{k}%
_{1}^{2}+2\upsilon ^{2})(\mathbf{k}_{1}^{2}-2G_{11}^{tr})}}),
\label{eq:integrated}
\end{gather}%
\newline
\newline
which shall be used for analytic continuation described below. Next, we
insert the $\Lambda _{mnl}(k)$ solved from Eq.\eqref{eq:self-con-k} into Eq.%
\eqref{eq:sol-gap-k}, so as to obtain the improved two-point vertices $%
\Gamma _{mn}(k)$. The improved two-point correlation functions take the form%
\begin{eqnarray}
G_{12}(k) &=&\frac{1}{M(k)}\left[ i\omega -\left( \mathbf{k}^{2}+\upsilon
^{2}-G_{11}^{tr}+\upsilon \left( \Lambda _{222}^{tr}(k)+2\Lambda
_{122}^{tr}(k)\right) \right) \right] ,  \notag \\
G_{22}(k) &=&\frac{1}{M(k)}\left[ \upsilon ^{2}+G_{11}^{tr}+\upsilon \left(
\Lambda _{221}^{tr}(k)+2\Lambda _{121}^{tr}(k)\right) \right] ,  \notag \\
G_{11}(k) &=&\frac{1}{M(k)}\left[ \upsilon ^{2}+G_{11}^{tr}+\upsilon \left(
\Lambda _{112}^{tr}(k)+2\Lambda _{122}^{tr}(k)\right) \right] ,  \notag \\
G_{21}(k) &=&\frac{1}{M(k)}\left[ -i\omega -\left( \mathbf{k}^{2}+\upsilon
^{2}-G_{11}^{tr}+\upsilon \left( \Lambda _{111}^{tr}(k)+2\Lambda
_{121}^{tr}(k)\right) \right) \right] ,  \label{IGA-G}
\end{eqnarray}%
where $M(k)$ is the determinant of the matrix $\Gamma _{mn}(k)$ and reads
\begin{gather}
M(k)=\left[ \upsilon ^{2}+G_{22}^{tr}+\upsilon \left( \Lambda
_{221}^{tr}(k)+2\Lambda _{121}^{tr}(k)\right) \right]  \notag \\
\times \left[ \upsilon ^{2}+G_{11}^{tr}+\upsilon \left( \Lambda
_{112}^{tr}(k)+2\Lambda _{122}^{tr}(k)\right) \right]  \notag \\
-\left[ -i\omega -\mu +\mathbf{k}^{2}+2\upsilon ^{2}+2G_{12}^{tr}+\upsilon
\left( \Lambda _{222}^{tr}(k)+2\Lambda _{122}^{tr}(k)\right) \right]  \notag
\\
\times \left[ i\omega -\mu +\mathbf{k}^{2}+2\upsilon
^{2}+2G_{12}^{tr}+\upsilon \left( \Lambda _{111}^{tr}(k)+2\Lambda
_{121}^{tr}(k)\right) \right] .
\end{gather}

The Green function in Eq. (\ref{IGA-G}) gives a gapless excitation spectrum,
which shall be shown by the numerical result and also can be analytically
verified by investigating the poles of the Green's function. Analytically,
one can prove $M(0)=0$ to make sure that the excitation spectrum is gapless.
The details of the proof are put in the Appendix.

One can obtain the real time Green's function, retarded and advanced Green's
function by analytic continuation, $i\omega \rightarrow \Omega \pm i\eta $,
where $\eta $ is an infinitesimal positive number. The spectral weight
function is then obtained by using the relation \cite{Fetter}
\begin{equation}
\rho \left( \mathbf{k},\Omega \right) =2\text{Im}G^{A}\left( \mathbf{k}%
,\Omega \right) =-2\text{Im}G^{R}\left( \mathbf{k},\Omega \right) .
\end{equation}%
Eq. (\ref{eq:integrated}) is an analytic function of "complex" variable $%
\Omega $ except on the real axis in $\Omega $ plane. The retarded and
advanced Green's function obtained therefore have desirable analytic
properties.

There is an equivalent formalism of the IGA approximation in the framework
of the improved $\Phi $ derivable theory. The $\Phi $ derivable theory can
start with the two particle irreducible (2PI) action functional $\widetilde{%
\Gamma }\left[ \varphi _{1},\varphi _{2},\mathbf{G}\right] $ which takes the
form

\begin{eqnarray}
\widetilde{\Gamma }\left[ \varphi _{1},\varphi _{2},\mathbf{G}\right] &=&S%
\left[ \varphi _{1},\varphi _{2}\right] +\frac{1}{2}Tr\ln \mathbf{G}^{-1}+%
\frac{1}{2}Tr\left[ \mathbf{D}^{-1}\left( \mathbf{G}-\mathbf{D}\right) %
\right] +\Phi \left[ \varphi _{1},\varphi _{2},\mathbf{G}\right]
\label{derivable} \\
\left( \mathbf{D}^{-1}\right) _{ij} &=&\frac{\delta ^{2}S\left[ \varphi
_{1},\varphi _{2}\right] }{\delta \varphi _{i}\delta \varphi _{j}},  \notag
\end{eqnarray}

where $(\varphi _{1},\varphi _{2})\equiv (\varphi ^{\ast },\varphi )$ as
defined previously, and $\mathbf{G}$ represents matrix $\left( \mathbf{G}%
\right) _{ij}=G_{ij}$ of Green's functions. In the order of HFB
approximation (omitting higher order diagrams like the setting sun diagram),
\begin{equation}
\Phi \left[ \varphi _{1},\varphi _{2},\mathbf{G}\right] =\frac{1}{2}\int dx%
\left[ G_{11}\left( x,x\right) G_{22}\left( x,x\right) +2G_{12}\left(
x,x\right) G_{21}\left( x,x\right) \right]
\end{equation}

We will obtain the same equations as Eq.(\ref{eq:HFB-shift}) and Eq.(\ref%
{eq:HFB-gap}) of the HFB approximation if we require%
\begin{equation}
\frac{\delta \widetilde{\Gamma }\left[ \varphi _{1},\varphi _{2},\mathbf{G}%
\right] }{\delta \varphi _{i}}=0,\frac{\delta \widetilde{\Gamma }\left[
\varphi _{1},\varphi _{2},\mathbf{G}\right] }{\delta G_{ij}}=0.
\label{shiftandgap}
\end{equation}%
In the framework of the $\Phi $ derivable theory, IGA can be reformulated as
below Ref.\cite{Hendrik}. The 1PI effective action $\Gamma \left[ \varphi
_{1},\varphi _{2}\right] $ is equal to $\widetilde{\Gamma }\left[ \varphi
_{1},\varphi _{2},\mathbf{G}^{tr}\left( \varphi _{1},\varphi _{2}\right) %
\right] $ with $\mathbf{G}^{tr}\left( \varphi _{1},\varphi _{2}\right) $
defined by $\frac{\delta \widetilde{\Gamma }\left[ \varphi _{1},\varphi _{2},%
\mathbf{G}\right] }{\delta G_{ij}}|_{\mathbf{G}=\mathbf{G}^{tr}\left(
\varphi _{1},\varphi _{2}\right) }=0$. Then from $\Gamma \left[ \varphi
_{1},\varphi _{2}\right] $, one obtains the inverse Green's function $\Gamma
_{ij}=\frac{\delta ^{2}\Gamma \left[ \varphi _{1},\varphi _{2}\right] }{%
\delta \varphi _{i}\delta \varphi _{j}}=\frac{\delta ^{2}\widetilde{\Gamma }%
\left[ \varphi _{1},\varphi _{2},\mathbf{G}^{tr}\left( \varphi _{1},\varphi
_{2}\right) \right] }{\delta \varphi _{i}\delta \varphi _{j}}$. For
technical details, see Ref.\cite{Hendrik}. Substituting the solution of Eq.(%
\ref{shiftandgap}) to the functional $\widetilde{\Gamma }\left[ \varphi
_{1},\varphi _{2},\mathbf{G}\right] $, we obtain a quantity $\Gamma $. The
thermodynamical potential is $\beta ^{-1}\Gamma $. According to the
thermodynamical relation, the particle number density $n$ is equal to $-\frac{%
\partial \Gamma }{\beta L\partial \mu }$ with $L$ being the size of the 1D system( $L$ is infinity in the
thermodynamic limit). Using Eq.(\ref{shiftandgap}) and
Eq.(\ref{derivable}), we know the density $n$ is equal to $\upsilon
^{2}+G_{12}^{tr}$, the same as the case of HFB. It is also valid in any $%
\Phi $ derivable theory or improved $\Phi $ derivable theory beyond HFB.

\section{Comparison with the Exact Solution}

The references \cite{key-19,key-29} present an exact solution of the
Lieb-Liniger model, which gives the exact excitation spectrum.

\begin{figure}[tbp]
\includegraphics[scale=0.9]{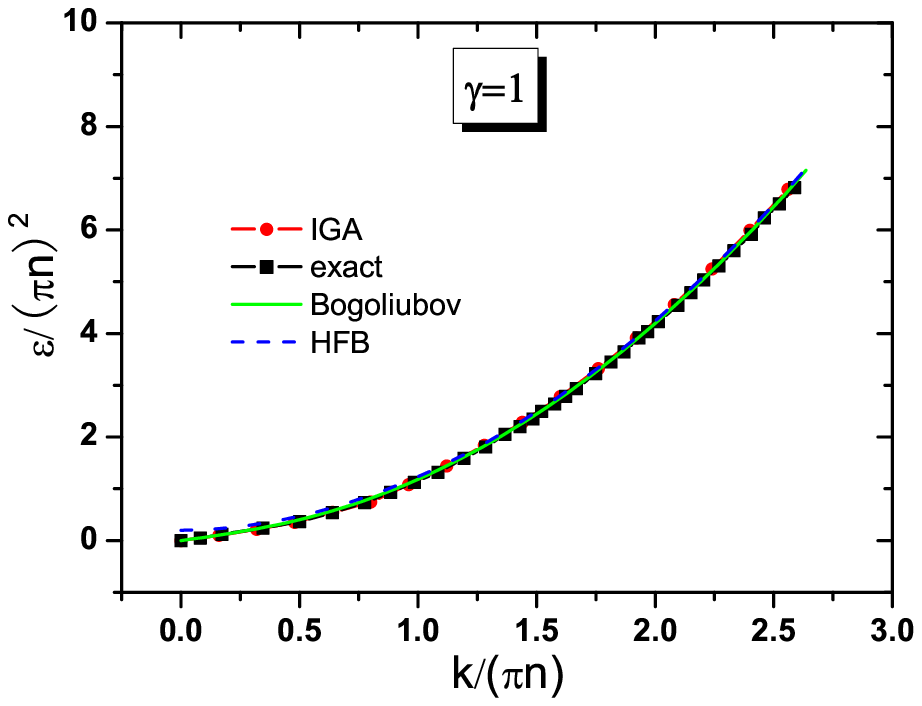}\\
\includegraphics[scale=0.9]{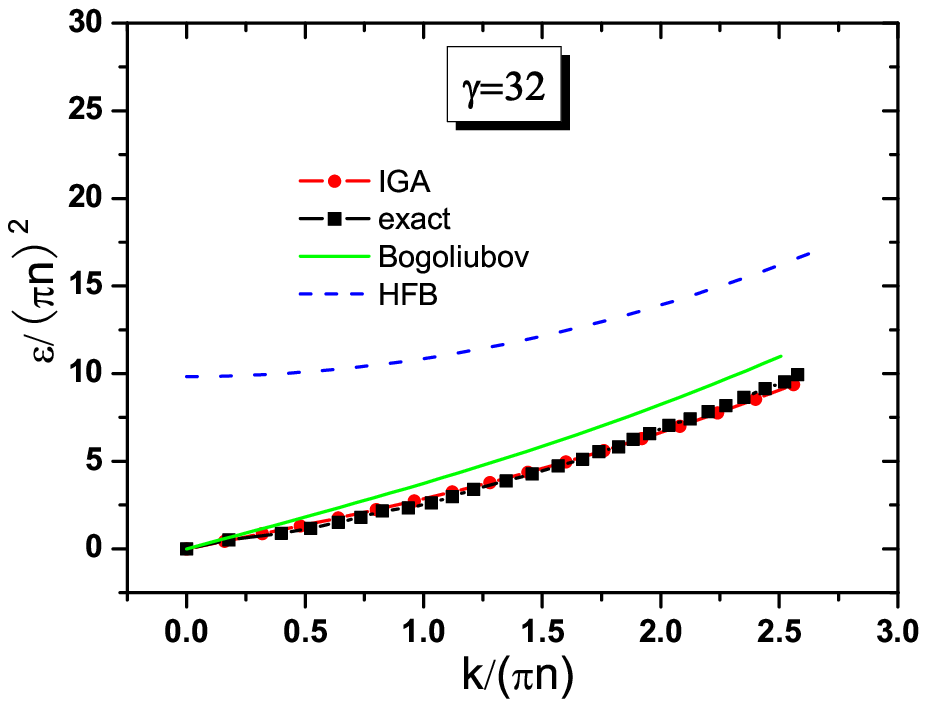}\\
\includegraphics[scale=0.9]{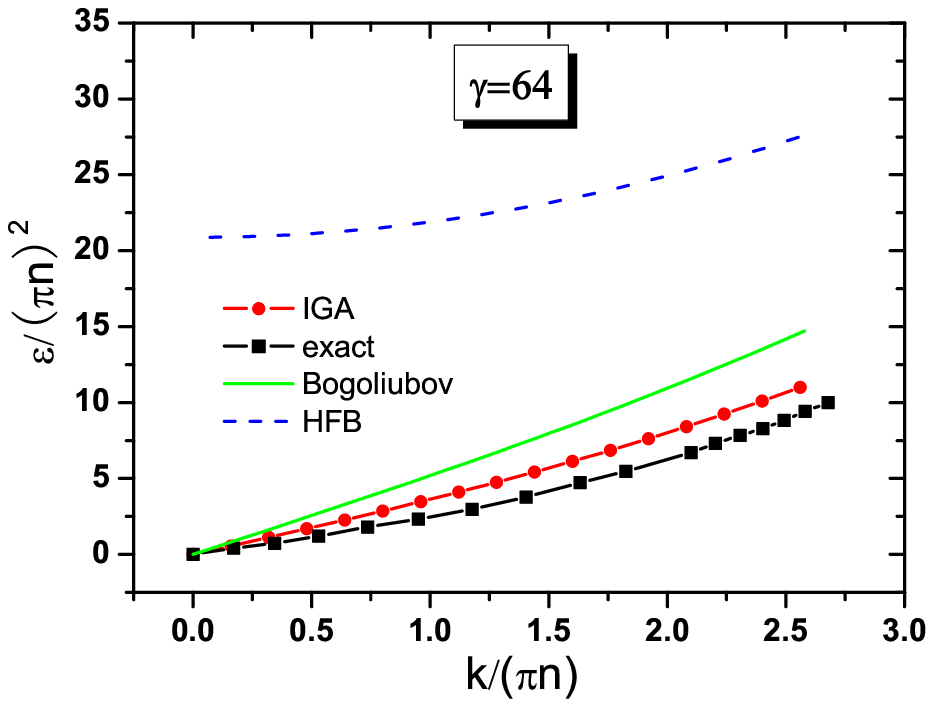}
\caption{(color online)The spectra obtained from Bogoliubov approximation
(green lines), HFB (blue dashed lines), IGA (red dotted lines) and exact
numerical calculations (black squares) from Bethe ansatz \protect\cite%
{key-29} are compared at three different $\protect\gamma $.}
\label{fig2}
\end{figure}

The references \cite{key-19,key-29} consider a one-dimensional system of
length $L$ (satisfying periodic boundary conditions), with $N$ bosonic
particles interacting via a repulsive contact potential of strength $2c$,
governed by the Lieb-Liniger Hamiltonian%
\begin{equation}
H=-\sum_{i=1}^{N}\left( \partial ^{2}/\partial x_{i}^{2}\right)
+2c\sum_{1\leq i<j\leq N}\delta (x_{i}-x_{j}).  \label{eq:LL}
\end{equation}%
The excitation spectrum is plotted as $\omega /n^{2}\sim k/n$ , with $n$
being the particle number density $\frac{N}{L}$. The dimensionless parameter
of the system is defined by $\gamma =\frac{c}{n}$. Comparing Eq.%
\eqref{eq:Hamiltonian} with Eq.\eqref{eq:LL}, there is $g=2c$ and hence the
corresponding parameter in the field-theoretic treatment takes the form $%
\gamma =\frac{g}{2n}$.Comparing Eqs.\eqref{eq:action-1}%
\eqref{eq:action-prime}, we know $n=gn^{\prime }$, $\mathbf{k}=g\mathbf{k}%
^{\prime }$ and $\omega =g^{2}\omega ^{\prime }$, which implies $\omega
/n^{2}=\omega ^{\prime }/n^{\prime 2}$, $\mathbf{k}/n=\mathbf{k}^{\prime
}/n^{\prime }$, where we restore the notation $\mathbf{k}^{\prime }$, $%
n^{\prime }$, $\omega ^{\prime }$ for the rescaled quantities after Eq.%
\eqref{eq:action-prime} (we had dropped prime for simplicity). Therefore, in
order to compare with the exact solution, we should plot the excitation
spectrum in the form $\omega ^{\prime }/n^{\prime 2}\sim k^{\prime
}/n^{\prime }$ with $n^{\prime }$ being the rescaled particle number
density, at the parameter $\gamma =\frac{1}{2n^{\prime }}$. At the
parameters $\gamma =1$, $\gamma =32$, $\gamma =64$, corresponding to $%
n^{\prime }=1/2$, $n^{\prime }=1/64$, $n^{\prime }=1/128$, we plot the
spectrum obtained from the different approximation schemes in Fig. \ref{fig2}%
. \ The IGA spectrum, which incorporates extra corrections based on the HFB
spectrum, is gapless, while the HFB spectrum is gapped. When the particle
density is high ($\gamma $ is small) , all approximation schemes lead to
good results, which implies that quantum fluctuations are weak at a high
particle density. Furthermore, at a very low particle density when quantum
fluctuations become strong, the IGA scheme shows its advantage.
Specifically, at $\gamma =32$ and $\gamma =64$, the IGA spectrum is in good
agreement with the exact one, while the Bogoliubov spectrum is not accurate
quantitatively.

\section{Summary}

We have presented IGA (improved Gaussian approximation) to treat one
dimensional bosonic gas. The Green's function obtained by IGA satisfies Ward
identities from $U(1)$ symmetry and therefore the spectrum is gapless.

We have formulated all the traditional approximations (Bogoliubov
approximation, HFB approximation and Popov approximation) in terms of
truncations of Dyson-Schwinger equations. The HFB approximation is the
well-known self-consistent approximation, but it leads to a gapped
excitation spectrum, violating the Goldstone theorem. The spectrum obtained
by IGA scheme, which incorporates more quantum corrections to the HFB
spectrum, is gapless. In order to test the validity and precision of the IGA
method, we apply it to the one-dimensional bosonic gas described by the
Lieb-Liniger model. We can only obtain Type I excitation ( Bogoliubov
spectrum) within IGA method. Nevertheless, by comparison with the type I
spectrum exactly solved by the Bethe ansatz, we find that the IGA method
gives quantitatively good results on type I spectrum.

The idea of the IGA method can be applied to improve higher order $\Phi $
derivable theory (the HFB theory is the result of the lowest order $\Phi $
derivable approximation) \cite{Hendrik,Rosenstein}. The essence of the idea
is to add extra Feynman diagrams to preserve the symmetry of all the Feynman
diagrams, and thereby restore the Ward identity. The IGA method makes
improvement based on the HFB approximation. When quantum fluctuations are
very strong, higher order $\Phi $ derivable approximation beyond the HFB
approximation will be required and then the corresponding improvement to
restore the Ward identity can be performed in a similar way. In order to get
type II (Fermionic excitation), one probably shall go beyond IGA and use
"improved" high order $\Phi $ derivable theory.

The IGA method presented here can be employed to handle many other
Bose-condensed systems, including 2D or 3D, zero temperature or finite
temperature, homogeneous or in optical lattices. For higher dimension
systems, as there is no type II excitation and quantum fluctuations are
weaker, IGA shall be expected to give more quantitatively accurate results.

As one of applications of IGA, we have carried out the IGA calculation on
type II superconductor where acoustic and optical spectra are obtained
non-perturbatively. The results will be presented elsewhere \cite{In
preparation}.

\begin{acknowledgments}
We thank Professor B. Rosenstein and Professor Zhongshui Ma for valuable
discussions. The work is supported by \textquotedblleft the Fundamental
Research Funds for the Central Universities\textquotedblright\ and National
Natural Science Foundation (Grant No. 10974001).
\end{acknowledgments}

\section{Appendix}

We shall prove $M(0)=0$, namely,
\begin{eqnarray}
0 &=&\left[ \upsilon ^{2}+G_{22}^{tr}+\upsilon \left( \Lambda
_{221}^{tr}(0)+2\Lambda _{121}^{tr}(0)\right) \right]   \notag \\
&&\times \left[ \upsilon ^{2}+G_{11}^{tr}+\upsilon \left( \Lambda
_{112}^{tr}(0)+2\Lambda _{122}^{tr}(0)\right) \right]   \notag \\
&&-\left[ -\mu +2\upsilon ^{2}+2G_{12}^{tr}+\upsilon \left( \Lambda
_{222}^{tr}(0)+2\Lambda _{122}^{tr}(0)\right) \right]   \notag \\
&&\times \left[ -\mu +2\upsilon ^{2}+2G_{12}^{tr}+\upsilon \left( \Lambda
_{111}^{tr}(0)+2\Lambda _{121}^{tr}(0)\right) \right] .  \label{eq:M(0)}
\end{eqnarray}%
By using the homogeneous and static condition in Eq.\eqref{eq:homo-condition}
and applying the Fourier transformation in Eq.\eqref{eq:FT}, we rewrite Eq.(%
\ref{eq:sol-shift}) as

\begin{eqnarray}
0 &=&-\mu +\upsilon ^{2}+G_{11}^{tr}+2G_{12}^{tr},  \notag \\
G_{11}^{tr} &=&G_{22}^{tr},  \label{eq:IGA-shift-tr}
\end{eqnarray}%
and Eq.(\ref{eq:gapeq}) as
\begin{eqnarray}
\Gamma _{11}^{tr}(k) &=&\upsilon ^{2}+G_{11}^{tr},  \notag \\
\Gamma _{22}^{tr}(k) &=&\upsilon ^{2}+G_{11}^{tr},  \notag \\
\Gamma _{12}^{tr}(k) &=&-i\omega +\mathbf{k}^{2}+\upsilon ^{2}-G_{11}^{tr},
\notag \\
\Gamma _{21}^{tr}(k) &=&i\omega +\mathbf{k}^{2}+\upsilon ^{2}-G_{11}^{tr}.
\label{eq:IGA-gap-tr}
\end{eqnarray}%
Note that the value of the external sources $J_{1}$ and $J_{2}$ has been set
to zero. With the help of Eq.(\ref{eq:G-Gamma-k}) we obtain the two-point
truncated Green's function,%
\begin{eqnarray}
\left(
\begin{array}{cc}
G_{11}^{tr}(k) & G_{12}^{tr}(k) \\
G_{21}^{tr}(k) & G_{22}^{tr}(k)%
\end{array}%
\right) &=&\frac{1}{(i\omega )^{2}-\left( \mathbf{k}^{2}+2\upsilon
^{2}\right) \left( \mathbf{k}^{2}-2G_{11}^{tr}\right) }  \notag \\
&&\times \left(
\begin{array}{cc}
\upsilon ^{2}+G_{11}^{tr} & i\omega -\left( \mathbf{k}^{2}+\upsilon
^{2}-G_{11}^{tr}\right) \\
-i\omega -\left( \mathbf{k}^{2}+\upsilon ^{2}-G_{11}^{tr}\right) & \upsilon
^{2}+G_{11}^{tr}%
\end{array}%
\right) ,  \label{eq:tr-G2}
\end{eqnarray}%
which has the same form as the Green's function in the HFB approximation.
Using Eq.\eqref{eq:IGA-shift-tr} we can rewrite Eq.\eqref{eq:M(0)} as
\begin{eqnarray}
0 &=&\left[ \upsilon ^{2}+G_{11}^{tr}+\upsilon \left( \Lambda
_{221}^{tr}(0)+2\Lambda _{121}^{tr}(0)\right) \right]  \notag \\
&&\times \left[ \upsilon ^{2}+G_{11}^{tr}+\upsilon \left( \Lambda
_{112}^{tr}(0)+2\Lambda _{122}^{tr}(0)\right) \right]  \notag \\
&&-\left[ \upsilon ^{2}-G_{11}^{tr}+\upsilon \left( \Lambda
_{222}^{tr}(0)+2\Lambda _{122}^{tr}(0)\right) \right]  \notag \\
&&\times \left[ \upsilon ^{2}-G_{11}^{tr}+\upsilon \left( \Lambda
_{111}^{tr}(0)+2\Lambda _{121}^{tr}(0)\right) \right] .  \label{eq:M(0)-1}
\end{eqnarray}%
By inserting Eq.\eqref{eq:tr-G2} in Eq.(\ref{eq:fish}), it is easy to verify
that
\begin{equation}
I_{ml,ln}(0)=I_{\bar{m}\bar{l},\bar{l}\bar{n}}(0)\,,  \label{I-Ibar}
\end{equation}%
where $m,n,l,\bar{m},\bar{n},\bar{l}=1,2$ with the constraint $\delta _{m%
\bar{m}}=0$, $\delta _{n\bar{n}}=0$ and $\delta _{l\bar{l}}=0$. For example,
$I_{11,11}(0)=I_{22,22}(0)$, $I_{21,11}(0)=I_{12,22}(0)$, etc. Eq.(\ref%
{I-Ibar}) and Eq.(\ref{eq:self-con-k}) lead to
\begin{equation}
\Lambda _{mnl}^{tr}(0)\equiv \Lambda _{\bar{m}\bar{n}\bar{l}}^{tr}(0)\,.
\label{eq:Lambda-Lambda-bar}
\end{equation}%
There is
\begin{equation}
\Lambda _{mnl}^{tr}(k)=\Lambda _{nml}^{tr}(k)\,,
\label{eq:Lambda_mnl-Lambda_nml}
\end{equation}%
which is evident from the definition in Eq.(\ref{Lambda-tr}). Using Eq.%
\eqref{eq:Lambda-Lambda-bar}and Eq.\eqref{eq:Lambda_mnl-Lambda_nml} we can
rewrite Eq.\eqref{eq:M(0)-1} as
\begin{eqnarray}
0 &=&\left( 2\upsilon +4\Lambda _{121}^{tr}(0)+\Lambda
_{111}^{tr}(0)+\Lambda _{221}^{tr}(0)\right)  \notag \\
&&\times \left( 2G_{11}^{tr}+\upsilon \Lambda _{221}^{tr}(0)-\upsilon
\Lambda _{111}^{tr}(0)\right) .  \label{eq:M(0)-2}
\end{eqnarray}%
From Eq.(\ref{eq:self-con-k}) we find that
\begin{equation}
\left( \Lambda _{221}^{tr}(0)-\Lambda _{111}^{tr}(0)\right)
=\frac{2\upsilon \left( I_{11,11}(0)-I_{12,21}(0)\right) }{\left(
1+I_{11,11}(0)-I_{12,21}(0)\right) }\,.  \label{eq:end1}
\end{equation}%
By straightforward calculations, we know
\begin{equation}
I_{11,11}(0)-I_{12,21}(0)=\frac{1}{4\pi }\int_{0}^{\infty }dk\frac{1}{\sqrt{%
\left( k^{2}+2\upsilon ^{2}\right) \left( k^{2}-2G_{11}^{tr}\right)
}}\,,\label{eq:I1111-I1221}
\end{equation}%
and

\begin{equation}
G_{11}^{tr}=-\left( \upsilon ^{2}+G_{11}^{tr}\right) \frac{1}{4\pi }%
\int_{0}^{\infty }dk\frac{1}{\sqrt{\left( k^{2}+2\upsilon ^{2}\right) \left(
k^{2}-2G_{11}^{tr}\right) }}.  \label{eq:IGA-G11}
\end{equation}%
Eq.\eqref{eq:IGA-G11} follows from the definition of $G_{11}^{tr}$, that is $%
G_{11}^{tr}=G_{11}^{tr}(x,x)=\int \frac{d\omega }{2\pi }\int \frac{d^{D}%
\mathbf{k}}{(2\pi )^{D}}G_{11}^{tr}(\mathbf{k},\omega )$. By comparing Eq.%
\eqref{eq:I1111-I1221} and Eq.\eqref{eq:IGA-G11} we know%
\begin{equation}
I_{11,11}(0)-I_{12,21}(0)=\frac{-G_{11}^{tr}}{\left( \upsilon
^{2}+G_{11}^{tr}\right) }\,.  \label{eq:end2}
\end{equation}%
Eq.\eqref{eq:end1} and Eq.\eqref{eq:end2} lead to
\begin{equation}
2G_{11}^{tr}+\upsilon \Lambda _{221}^{tr}(0)-\upsilon \Lambda
_{111}^{tr}(0)=0.
\end{equation}%
Thus Eq.\eqref{eq:M(0)-2} is proved and also Eq.\eqref{eq:M(0)} is proved.


\begin{thebibliography}{99}
\bibitem{key-1} N. N. Bogoliubov, J. Phys. (Moscow) \textbf{11}, 23 (1947).

\bibitem{key-2} N. N. Bogoliubov, Moscow Univ. Phys. Bull. 7, 43 (1947).

\bibitem{key-3} N. N. Bogoliubov, \textit{Lectures on Quantum Statistics}
(Gordon and Breach, New York, 1967),Vol.1.

\bibitem{key-4} N. N. Bogoliubov, \textit{Lectures on Quantum Statistics}
(Gordon and Breach, New York, 1970),Vol.2.

\bibitem{key-5} S.T. Beliaev, J. Exp. Theor. Phys. \textbf{7}, 289 (1958).

\bibitem{key-6} S.T. Beliaev, J. Exp. Theor. Phys. 7, 299 (1958).

\bibitem{key-7} N. M. Hugenholtz and D. Pines, Phys. Rev. \textbf{116}, 489
(1959).

\bibitem{key-41} J. Goldstone, Nuovo Cimento \textbf{19}, 154 (1961).

\bibitem{key-8} P. C. Hohenberg and P. C. Martin, Ann. Phys. \textbf{34},
291 (1965).

\bibitem{key-33} J. Luttinger and J. Ward, Phys. Rev. \textbf{118}, 1417
(1960).

\bibitem{key-34} C. de Dominicis and P.C. Martin, J. Math. Phys. \textbf{5},
14 (1964).

\bibitem{key-35} C. de Dominicis and P.C. Martin, J. Math. Phys. \textbf{5},
31 (1964).

\bibitem{key-36} J. M. Cornwall, R. Jackiw, and E. Tom boulis, Phys. Rev. D
\textbf{10}, 2428 (1974).

\bibitem{Baym} G. Baym and Leo P. Kadanoff , Phys. Rev. \textbf{124}, 287
(1961). G. Baym, Phys. Rev. \textbf{127}, 1391 (1962).

\bibitem{key-9} A. Griffin, Phys. Rev. B \textbf{53}, 9341 (1996).

\bibitem{key-10} H. Shi and A. Griffin, Phys. Rep. \textbf{304}, 1 (1998).

\bibitem{key-11} D. A. W. Hutchinson \textit{et al.}, J. Phys. B \textbf{33}%
, 3825 (2000).

\bibitem{key-12} V. N. Popov, \textit{Functional Integrals in Quantum Field
Theory and Statistical Physics} (Reidel, Dordrecht, 1983).

\bibitem{key-13} E. Lundh and J. Rammer, Phys. Rev. A \textbf{66}, 033607
(2002).

\bibitem{key-14} J. O. Andersen, Rev. Mod. Phys. \textbf{76}, 599 (2004).

\bibitem{Hendrik} H. van Hees and J. Knoll, Phys. Rev. D \textbf{66}, 025028
(2002).

\bibitem{Rosenstein} A. Kovner and B. Rosenstein, Phys. Rev. D \textbf{39},
2332 (1989). B. Rosenstein and A. Kovner, Phys. Rev. D \textbf{40}, 504
(1989).

\bibitem{Okopinska} A. Okopi\'{n}ska, Physics Letters B, \textbf{375}, 213
(1996). A. Okopi\'{n}ska, arXiv:cond-mat/0309679v1.

\bibitem{key-22} A. G\"{o}rlitz \textit{et al}., Phys. Rev. Lett. \textbf{87}%
, 130402 (2001).

\bibitem{key-23} H. Moritz, T. St\"{o}ferle, M. K\"{o}hl, and T. Esslinger,
Phys. Rev. Lett. \textbf{91}, 250402 (2003).

\bibitem{key-24} B. Paredes\textit{\ et al}., Nature (London) \textbf{429},
277 (2004).

\bibitem{key-25} T. Kinoshita, T. Wenger, and D. S. Weiss, Science \textbf{%
305}, 1125 (2004).

\bibitem{ShlyapnikovPRL2000} D.S. Petrov, G.V. Shlyapnikov, and J.T.M.
Walraven, Phys. Rev. Lett. \textbf{85}, 3745 (2000).

\bibitem{ShlyapnikovPRL2003} D.M. Gangardt and G.V. Shlyapnikov, Phys. Rev.
Lett. \textbf{90}, 010401 (2003).

\bibitem{key-26} M. A. Cazalilla \textit{et al.}, Rev. Mod. Phys. \textbf{83}%
, 1405 (2011), references therein.

\bibitem{key-19} E. Lieb, Phys. Rev. \textbf{130}, 1616 (1963).

\bibitem{key-20} E. Lieb and W. Liniger, Phys. Rev. \textbf{130}, 1605
(1963).

\bibitem{key-27} M. Olshanii, Phys. Rev. Lett. \textbf{81}, 938 (1998).

\bibitem{key-28} V. Dunjko, V. Lorent, and M. Olshanii, Phys. Rev. Lett.
\textbf{86}, 5413 (2001).

\bibitem{key-29} J. S. Caux, P. Calabrese and N. A. Slavnov, Journal of
Statistical Mechanics: Theory and Experiment \textbf{2007}, P01008 (2007);
J. S. Caux and P. Calabrese, Phys. Rev. A \textbf{74}, 031605(R) (2006).

\bibitem{key-30} Algebraic Bethe Ansatz Computation of Universal Structure
factors. See http://staff.science.uva.nl/jcaux/ABACUS.html.

\bibitem{key-38} J. W. Negele and H. Orland, \textit{Quantum Many-Particle
Systems} (Westview Press, 1988).

\bibitem{key-39} R. Jackiw, Phys. Rev. D \textbf{9}, 1686 (1974).

\bibitem{Stoof} J.O. Andersen, U. Al Khawaja, and H.T.C. Stoof, Phys. Rev.
Lett. \textbf{88}, 070407 (2002);U. Al Khawaja, J.O. Andersen, N.P.
Proukakis, and H.T.C Stoof, Phys. Rev. \textbf{A66} 013615 (2002).

\bibitem{key-32} E. Calzetta and B. L. Hu, \textit{Nonequilibrium quantum
field theory}, Chap.13 (Cambridge University Press, 2008).

\bibitem{Fetter} A. L. Fetter and J. D. Walecka, \textit{Quantum Theory of
Many Particle Systems} (McGraw-Hill, New York, 1971).

\bibitem{In preparation} Qiong Li, Li Zhang\quad and Dingping Li, in
preparation.
\end{thebibliography}
\end{document}